\documentstyle[prb,aps,epsf,floats]{revtex}
\begin{document}
\renewcommand{\textfraction}{0.0}
\renewcommand{\floatpagefraction}{.7}
\setcounter{topnumber}{5}
\renewcommand{\topfraction}{1.0}
\setcounter{bottomnumber}{5}
\renewcommand{\bottomfraction}{1.0}
\setcounter{totalnumber}{5}
\setcounter{dbltopnumber}{2}
\renewcommand{\dbltopfraction}{0.9}
\renewcommand{\dblfloatpagefraction}{.7}
\draft

\title{\huge{\bf Generalization in the Hopfield Model}}
\vskip 10mm
\author{\huge Leonid B. Litinskii}
\vskip 10mm
\address{\huge High Pressure Physics Institute \\of Russian Academy of 
Sciences\\
{\it Russia, 142092 Troitsk Moscow region,\\ e-mail: u10740@dialup.podolsk.ru}}

\maketitle
\vskip 10mm
\begin{abstract}
\LARGE 
In the Hopfield model the ability of the network to generalization is studied
in the case of the network trained by one input image ({\it the standard}).
\end{abstract}
\vskip 15mm
\LARGE
\centerline{{\huge{\bf\underline{Basic Model}}}}
\vskip 3mm
\noindent
The maximization problem with the symmetrical connection
matrix is considered:
$$\left\{\begin{array}{l} F(\vec\sigma)=\sum\limits^n J_{ij}\sigma_i\sigma_j
\to\max, \ \sigma_i=\{\pm 1\} \ \forall i\\
J_{ij}=J_{ji},\mbox{ $J_{ii}$ \underline{\rm does not matter \cite{Lit3}}}.
\end{array}\right.\eqno(1)$$
The configuration vector $\vec\sigma=(\sigma_1,\sigma_2,\ldots,\sigma_n)$ 
providing the solution of the problem is called \underline{the ground state}.

\noindent
The Hebbian-like representation exists for every $\bf J$,
$$\mbox{\LARGE ${\bf J}={\bf S}^{\rm T}\cdot{\bf S}$},\quad
\mbox{where $\bf S$ is a real $(p\times n)$-matrix and $p=${\rm rank}\ $\bf J$.}
\eqno(2)$$
We would like to investigate the special case of the $(p\times n)$-matrix 
$\bf S$:
$$\bf S{\LARGE =\left(\begin{array}{ccccccc}
1-x&1&\ldots&1&1&\ldots&1\\
1&1-x&\ldots&1&1&\ldots&1\\
\vdots&\vdots&\ddots&\vdots&\vdots&\ldots&\vdots\\
1&1&\ldots&1-x&1&\ldots&1\end{array}\right),\quad\left\{\begin{array}{l}
x\mbox{ is real}\\p+q=n.\end{array}\right.} 
\eqno(3)$$
\vskip 2mm
\noindent
The rows of the matrix $\bf S$ are the {\it generalized} memorized patterns. 

\noindent
The meaningful interpretation of the problem:
\vskip 2mm
\noindent
\fbox{%
\parbox{175mm}{%
the network had to be learned by $p$-time showing of \underline{the
standard} $\vec\varepsilon (n)$, 
$$\vec\varepsilon(n)=(1,1,\ldots,1)\in {\rm R}^n,$$
but an error crept into the learning process and
the network was learned by $p$ distorted copies 
$\vec s^{(l)}$ of the standard:
$$\vec s^{(l)}=(1,\ldots,1,\underbrace{1-x}_l,1,\ldots,1),\quad l=1,2,\ldots,
p.\eqno(4)$$
}%
}
\vskip 2mm
\noindent
The real number $x$ is called \underline{the distortion parameter}.

\noindent The problem under investigation, Eqs.(1) -(3), is very close to the
problem of generalization in the case of one embedded pattern \cite{Fon,Kre}. 
\vskip 5mm
\centerline{{\huge{\bf\underline{Main Results}}}\cite{Lit5}}
\vskip 3mm
\noindent
${\bf 1^\circ.}$ The local maxima of the functional 
$F(\vec\sigma)$ necessarily have the form \cite{Lit1}
$$\vec\sigma^*=(\underbrace{\sigma_1,\sigma_2,\ldots,\sigma_p}_{\vec\sigma'},
1,\ldots,1),\eqno(5)$$
and
$$F(\vec \sigma^*)\propto x^2 - 2x(q+p\cos w)\cos w +(q+p\cos w)^2,\eqno(6)$$
where $w$ is the angle between vectors  
$\vec\sigma'$ and $\vec\varepsilon(p)=(1,1,\ldots,1)\in {\rm R}^p$:
$$\cos w=\frac{\sum_{i=1}^p \sigma_i}p=\frac{(\vec\sigma',\vec\varepsilon 
(p))}{\parallel\vec\sigma'\parallel\cdot\parallel\vec\varepsilon 
(p)\parallel}.
\eqno(7)$$
Then, the vectors $\vec\sigma^*$ (5) with the $p$-dimensional parts
$\vec\sigma'$ equidistant from $\vec\varepsilon (p)$, {\it provide the
same value} of the functional $F$.  
\vskip 3mm
\noindent
${\bf 2^\circ.}$ Evidently, 
$$\cos w\equiv\cos w_k=1-\frac {2k}p,\ k=0,1,\ldots,p,$$
and the vectors $\vec\sigma^*$ (5) are
grouped into $p+1$ {\it classes} $\Sigma_k$ on which the functional
$F(\vec\sigma^*)$ is constant:
$$\mbox{\fbox{$\Sigma_k=\{\vec\sigma^*\mid$\mbox{ 
exactly $k$ coordinates of $\vec\sigma^*$ are equal to "--1"\}.}}}$$ 
The number of the vectors $\vec\sigma^*$ in the class $\Sigma_k$ is equal
${p\choose k}$.
\vskip 3mm
\noindent
${\bf 3^\circ.}$  To find the ground state dependence on  
$x$, it is necessary to analyze the family of the straight lines
$$L_k(x)=(q+p\cos w_k)^2 - 2x(q+p\cos w_k)\cos w_k,\ k=0,1,\ldots,p. 
\eqno(8)$$
In the region where $L_k(x)$ majorizes all the other straight lines, the ground
state belongs to the class $\Sigma_k$ and it is ${p\choose k}$ times degenerated.
\vskip 3mm
\noindent
${\bf 4^\circ.{\mbox{\underline{Theorem.}}}}$
\vskip 2mm
\noindent
{\it 
\noindent
When $x$ increases from $-\infty$ to $\infty$, the ground state in consecutive
order belongs to the classes  
$$\Sigma_0,\Sigma_1,\ldots,\Sigma_{k_{max}}.$$
The $k${\rm th} rebuilding of the ground state ($\Sigma_{k-1}\to\Sigma_k$)
occurs at the point $x_k$ of the intersection of the straight lines 
$L_{k-1}(x)$ and $L_k(x)$:
$$x_k=p\frac{n-(2k-1)}{n+p-2(2k-1)},\quad k=1,2,\ldots,k_{max},\eqno(9)$$
where
$$k_{max}= \min\left(p,\left[\frac{n+p+2}4\right]\right).$$
The functional has no other local maxima.
}
\vskip 3mm
\noindent
The Theorem relates the quality of the learning of the network with the value
of the distortion $x$ during the learning stage and with the length $p$ of the
learning sequence. It is reasonable, that the error of the network increases
with the increase of the distortion $x$: when $x\in(x_k,x_{k+1})$ the class
$\Sigma_k$ ("the truth" understood by the network) differs from the standard
$\vec\varepsilon (n)$ by $k$ coordinates (others interpretations see
below). 
\newpage
\LARGE
\begin{figure}[htb]
\begin{center}
\leavevmode
\epsfxsize = 16.2truecm
\epsfysize = 8.5truecm
\epsffile{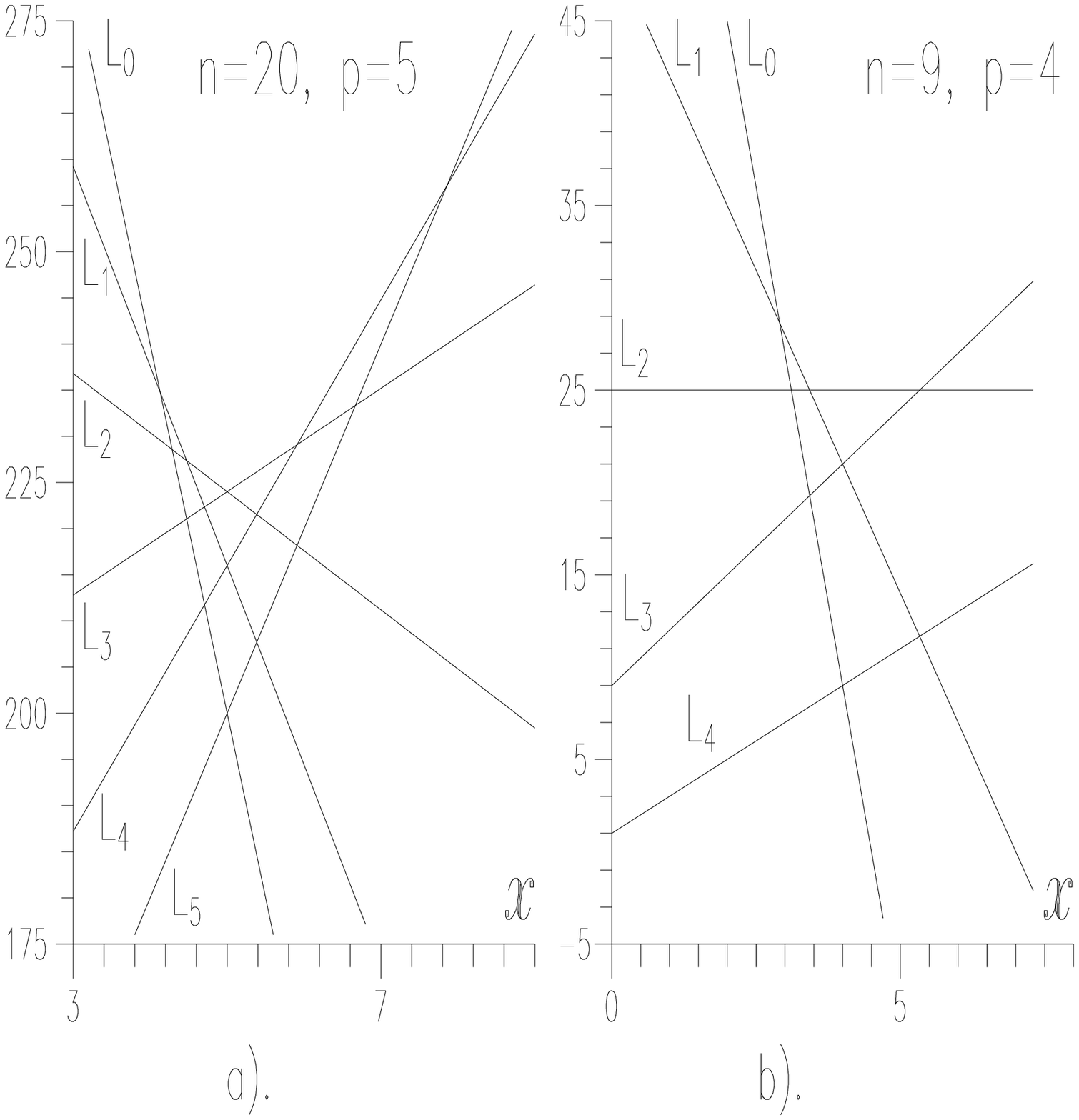}
\end{center}
\end{figure}
\noindent
In the Fig.1 is given the typical behavior of the straight lines 
$L_k(x)$. The rebuildings of the ground state occurs at the
points $x_k$ of the intersection of the straight lines $L_{k-1}$ and $L_k$.
Inside the interval $(x_k,x_{k+1})$ the ground state belongs to the class
$\Sigma_k$. When $x$ increases: 

\noindent
a). all the rebuildings of the ground state occur: $k_{max}=p$; 

\noindent
b). only 
$k_{max}=\left[\frac{n+p+2}4\right]<p$ rebuildings of 
the ground state occur.

\vskip 5mm
\centerline{{\huge{\bf\underline{Generalizations of Basic Model}}}\cite{Lit5}}
\vskip 3mm
\noindent
${\bf 1^\circ.}$  When the standard $\vec\varepsilon
(n)$ is changed by an {\it arbitrary} configuration vector  
\mbox{$\vec{\alpha}=(\alpha_1,\alpha_2,\ldots,\alpha_n),\ \alpha_i=\{\pm 1\}$},
the Theorem remains valid, but the vectors $\vec\sigma^*$ (5) have the form
$$\vec\sigma^*=(\alpha_1\sigma_1,\alpha_2\sigma_2,\ldots,\alpha_p\sigma_p,
\alpha_{p+1},\ldots,\alpha_n).$$
${\bf 2^\circ.}$ When we rotate the memorized patterns (4) as a whole, 
all their first $p$ coordinates are distorted.

\noindent
Suppose the rotation matrix ${\bf U}=(u_{ij})$ transforms
the first $p$ coordinates of $n$-dimensional vectors only:
$$\left\{\begin{array}{l}\vec u={\bf U}\cdot\vec\varepsilon
(p)=(u_1,u_2,\ldots,u_p),\\
u_l=\sum_{i=1}^p u_{li},\ l=1,2,\ldots,p;\quad\parallel\vec 
u\parallel^2=p.
\end{array}\right.\eqno(10)$$
Then the memorized patterns take the form:  
$$\vec s^{(l)}=(u_1-xu_{1l},u_2-xu_{2l},\ldots,u_p-xu_{pl},1,\ldots,1),
\quad l=1,2,\ldots,p.$$
\vskip 2mm
\noindent
It is easy to see, that if the standard $\vec\varepsilon (n)$ does not change
after the rotation ($u_l\equiv 1$) all the results of the "Basic
Model" \underline {remain unchanged}. 

\noindent
More interesting is the case when the standard $\vec\varepsilon (n)$ shifts
after the rotation ($u_l\ne 1$). Again the only important
configuration vectors are $\vec\sigma^*$ (5) and the functional
$F(\vec\sigma^*)$ is given by the same expression (6). But now $w$ is the angle
between vectors $\vec\sigma'$ and $\vec u$:  
$$\cos  w=\frac{\sum_{i=1}^p \sigma_i\cdot u_i}p=
\frac{(\vec\sigma',\vec u)}{\parallel \vec \sigma' 
\parallel \cdot \parallel \vec u \parallel}.\eqno(11)$$
The vectors $\vec\sigma^*$ are grouped in the classes $\Sigma_k^{(U)}$ 
on which $F(\vec\sigma^*)$ is constant:
$$\mbox{\fbox{$\Sigma_k^{(U)}=\{\vec\sigma^*\mid$ \ with 
$p$-dimensional parts $\vec\sigma'$ equidistant from $\vec u$.\}}}$$
The number of the different classes $\Sigma^{(U)}_k$ is given by the number $t$
of {\it different} values of $\cos w$ (11):   
$$\cos w_0>\cos w_1>\ldots>\cos w_t;\quad \cos w_k=-\cos w_{t-k},\quad\forall
k\le t.$$ 
Then we have the following generalization of the Theorem:
\vskip 2mm
\noindent
{\it when $x$ increases from $-\infty$ to $\infty$, the ground state 
in consecutive order belongs to the classes 
$$\Sigma^{(U)}_0,\Sigma^{(U)}_1,\ldots,\Sigma^{(U)}_{k_{max}}.$$
The $k${\rm th} rebuilding of the ground state 
($\Sigma^{(U)}_{k-1}\to\Sigma^{(U)}_k$) occurs at the point $x_k$ of the
intersection of the straight lines $L_{k-1}(x)$ and $L_k(x)$:
$$x_k=\frac{p}{2}\left[1+\frac{q}{q+p(\cos w_{k-1}+\cos w_k)}\right],
\quad k=1,2,\ldots,k_{max}.
\eqno(12)$$ 
If $x_1>\frac34p$, all the rebuildings take place ($k_{max}=t$). If
$x_1<\frac34p$, the rebuildings stop when the denominator in Eq.(12) becomes
negative.} 
\vskip 3mm
\noindent
{\bf Note.} The compositions of the classes $\Sigma_k^{(U)}$ are determined
by the values of $\{u_l\}_1^p$ only. But the choice of $\{u_l\}$ is
completely in our hands! Then we can create the Hopfield type
network with a preassigned set of the fixed points.

\noindent
${\bf 3^\circ.}$ The memorized patterns can be obtained from $\vec\varepsilon
(n)$ by {\it the identical synchronous} distortions of its $m$ coordinates.
Suppose $n=p\times m+q$ and the matrix ${\bf S}$ is
$${\bf S}=
\left(\begin{array}{ccccccccccccc}
1-x&\ldots&1-x&1&\ldots&1&\ldots&1&\ldots&1&1&\ldots&1\\
1&\ldots&1&1-x&\ldots&1-x&\ldots&1&\ldots&1&1&\ldots&1\\
\vdots&\ldots&\vdots&\vdots&\ldots&\vdots&\ldots&\vdots&\ldots&\vdots&\vdots&
\ldots&\vdots\\
1&\ldots&1&1&\ldots&1&\ldots&1-x&\ldots&1-x&1&\ldots&1\end{array}
\right).$$
\vskip 4mm
\noindent
The "suspicious" configuration vectors are the piecewise constant vectors 
$$\vec\sigma^*=(\underbrace{\sigma_1,\ldots,\sigma_1}_m,\underbrace{
\sigma_2,\ldots,\sigma_2}_m,\ldots,
\underbrace{\sigma_p,\ldots,\sigma_p}_m,\underbrace{1,\ldots,1}_q);\eqno(13)$$
the values of the functional are
$$F(\vec\sigma^*)\propto\ x^2 - 2x\cos w \left(\frac{q}m+p\cos w\right)+
\left(\frac{q}m+p\cos w\right)^2, $$
where, as in Eq.(7), $w$ is the angle between  
$\vec \sigma'=(\sigma_1,\sigma_2,\ldots,\sigma_p)$ and $\vec \varepsilon (p)$.
\vskip 2mm
\noindent
Again the vectors (13) are grouped into classes $\Sigma_k^{(m)}$, whose 
structure is similar to the structure of the classes $\Sigma_k$. Then we have
the generalization of the Theorem:  
{\it the value of the parameter $x$, which corresponds to the $k${\rm th}
rebuilding of the ground state ($\Sigma_{k-1}^{(m)}\to\Sigma_k^{(m)}$), is
$$x_k=p\frac{\frac{n}m-(2k-1)}{\frac{n}m+p-2(2k-1)},$$
and}
$$k_{max}= \min\left(p,\left[\frac{\frac{n}m+p+2}4\right]\right).$$

\newpage
\LARGE
\centerline{{\huge{\bf\underline{Basic Model: Sequences and
Interpretations}}}\cite{Lit5}} 
\vskip 10mm
\noindent
\mbox{\fbox{${\bf 1^\circ.}\quad x_1=p\frac{n-1}{n+p-2}\ge\frac{p}2.$}}
\vskip 7mm
\noindent
In fact $x_1$ is the boundary of the distortions up to which the network
reproduces the standard from its distorted copies correctly. The boundary
depends on the length $p$ of the learning sequence. Of course, the network is
learned correctly, if the value of the distortions does not exceed $\frac{p}2$.
\vskip 10mm
\noindent
\mbox{\fbox{${\bf 2^\circ.}\quad x_1 \mbox{ is monotonically increasing
function of } p 
\mbox{ and } n.$}}
\vskip 5mm
\noindent
\underline{\it Let $n$ and $x$ be fixed.} Merely due to an increase of $p$ the
boundary $x_1$ can be forced to exceed $x$ (if $x$ is not too 
large). As a result $x$ turns out to be on the left of a new position of $x_1$,
i.e. in the region where {\it the only} fixed point is the standard $\vec\varepsilon
(n)$. In other words, only by an increase of the length $p$ of the learning
sequence we can force the network to understand correctly "the truth"
it is tried to be learned. It is in agreement with the practical experience: 
the greater the length of the learning sequence, the better the signal can be
read through noise.  
\vskip 12mm
\noindent
\underline{\it Let $p$ and $x$ be fixed.} As above, merely due to an increase of the 
number $n$ the value of $x_1$ can be forced to exceed $x$. This result is
reasonable too: if $p$ is fixed, the greater is $n$, the smaller is the
relative weight $\frac{p}n$ of the distorted coordinates. Naturally, the less
is the relative distortion, the better must be the result of the learning.
\newpage
\LARGE
\noindent
\mbox{\fbox{${\bf 3^\circ.}\quad x_{\frac{p+1}2}=p$.}} (Without the loss of
generality we assume that $p$ is odd.)
\vskip 5mm
\noindent
Here we change the notation for the standard $\vec\varepsilon(n)$ and
introduce another standard $\vec\varepsilon^{(-)}$:
$$\begin{array}{crrcrrcl}\vec\varepsilon^{(+)}=(&1,&1,&\ldots,&1,&1,&\ldots,&1)
=\vec\varepsilon(n)\\
\vec\varepsilon^{(-)}=(&-1,&-1,&\ldots,&-1,&1,&\ldots,&1).\end{array}$$
\vskip 2mm
\noindent
In their not coincident parts the standards $\vec\varepsilon^{(+)}$ and
$\vec\varepsilon^{(-)}$ are opposed with each other, i.e. they are two opposite
"statements". Any of the network fixed points $\vec\sigma^*$ (5) is an
intermediate statement between $\vec\varepsilon^{(+)}$ and
$\vec\varepsilon^{(-)}$, which is drawn towards either one edge of the scale,
or the other. And the network "feels" this.

Indeed, when the distortion $x$ is not very large ($x<p$), the number $k$ of
the ground state does not exceed $\frac{p}2$, and the ground state more
resembles $\vec\varepsilon^{(+)}$ than $\vec\varepsilon^{(-)}$. In other words,
the memorized patterns are interpreted by the network as the distorted copies
of the standard $\vec\varepsilon^{(+)}$. But if during the learning stage the
distortion exceeds $p$ $(x>p)$, the number of ground
state exceeds $\frac{p}2$ and the ground state resembles
$\vec\varepsilon^{(-)}$. Now the network interprets the memorized patterns as
the distorted copies of another standard $\vec\varepsilon^{(-)}$. 

This is in agreement with the practical experience: we interpret deviations in
the image of a standard as permissible {\it only up to some boundary}. If only
this boundary is exceeded, the patterns are interpreted as the distortions of
quite different standard. For the network of the considered type this boundary
is $p$.  

One extra argument to support this interpretation:
from Eq.(9) it is easy to see that when $p=const$ and $n\to\infty$ all 
$x_k$ stick to one point 
$$x_k\equiv p,\quad k=1,2,3,\ldots,p;$$
then for $x<p$ the ground state belongs to the class
$\Sigma_0=\vec\varepsilon^{(+)}$, whereas for $x>p$ the ground state belongs to
the class $\Sigma_p=\vec\varepsilon^{(-)}$. 
\newpage
\LARGE
\noindent
\mbox{\fbox{${\bf 4^\circ.}\quad k_{max} =\left\{\begin{array}{cl}p,&\mbox{ 
when }\frac{p-1}{n-1}<\frac13\\
\left[\frac{n+p+2}4\right],&\mbox{ when }\frac{p-1}{n-1}>\frac13
\end{array}\right.$}}
\vskip 2mm
\noindent
So, $p$ is the boundary for the permissible distortions $x$. 
The question is, what do the memorized patterns 
with large distortions $x$ mean? We treat the increase of $x$ above 
$p$ as
the more and more \underline{negation} of the standard $\vec\varepsilon^{(+)}$.
As if the network is learned by the memorized patterns, which \underline{deny}
the standard $\vec\varepsilon^{(+)}$. In other words, the network is 
{\it relearned by presentation of negative examples}.

There is big and clear to everybody difference between the 
relearning with the help of negative examples and {\it the learning of
the opposite truth}. 
The relearning is characterized by some specific difficulties:
(1) the better the
incorrect truth has been understood, the more difficult (and sometimes even 
impossible) to correct it; (2) it is comparatively easy to correct the result
slightly, but it is much more difficult to revise it in the main, {\it etc.}
We think, that the dependence of $k_{max}$ on $p$ is the reflection of just
these problems.  

When the number $p$ of the parameters which
have to be corrected is not very great ($\frac{p-1}{n-1}<\frac13$), the
network can be relearned by simple presentation of negative examples. In this
case $k_{max}=p$ and, when "the denial" of the standard 
$\vec\varepsilon^{(+)}$
is rather strong ($x>x_p$), as "a new" truth the network understands
the opposite standard $\vec\varepsilon^{(-)}$. But if the number of the
corrected parameters is great ($\frac{p-1}{n-1}>\frac13$), to relearn the
network it is not sufficient to present the negative examples. In this case
$\left[\frac{p+1}2\right]<k_{max}<p$
and whatever large $x$ is $(x_{k_{max}}<x<\infty)$, as a new truth the network
understands not the opposite standard $\vec\varepsilon^{(-)}$, but one of the
statements intermediate between $\vec\varepsilon^{(+)}$ and 
$\vec\varepsilon^{(-)}$. Though the understood truth is drawn towards 
$\vec\varepsilon^{(-)}$, since $k_{max}>\frac{p}2$.

Of course, our interpretation is open for discussion. But it seems that in
real life there are a lot of examples, which confirm our conception.
\newpage
\LARGE
\centerline{{\bf\underline{\huge{New Results (in preparation)}}}}
\vskip 3mm
\noindent
The generalization of the Basic Model to the cases:

\noindent
${\bf 1^\circ.}$ The functional
$F(\vec\sigma)$ in the problem (1)-(3) has the form
$$F(\vec\sigma)=\sum\limits^n_{i,j=1} J_{ij}\sigma_i\sigma_j + h\sum_{i=1}^n\sigma_i.$$
In physics such a linear term describes the magnetic field. 

\noindent
In the Fig.2 the straight lines $h_k(x)$ divide the plane $(x,h)$ into the
regions where the ground state belongs to the different classes $\Sigma_k$, 
$$h_{k-1}(x)=2p(n-2k+1)\left(\frac{x}{x_k}-1\right),\ k=1,2,\ldots$$

\begin{figure}[htb]
\begin{center}
\leavevmode
\epsfxsize = 15truecm
\epsfysize = 10truecm
\epsffile{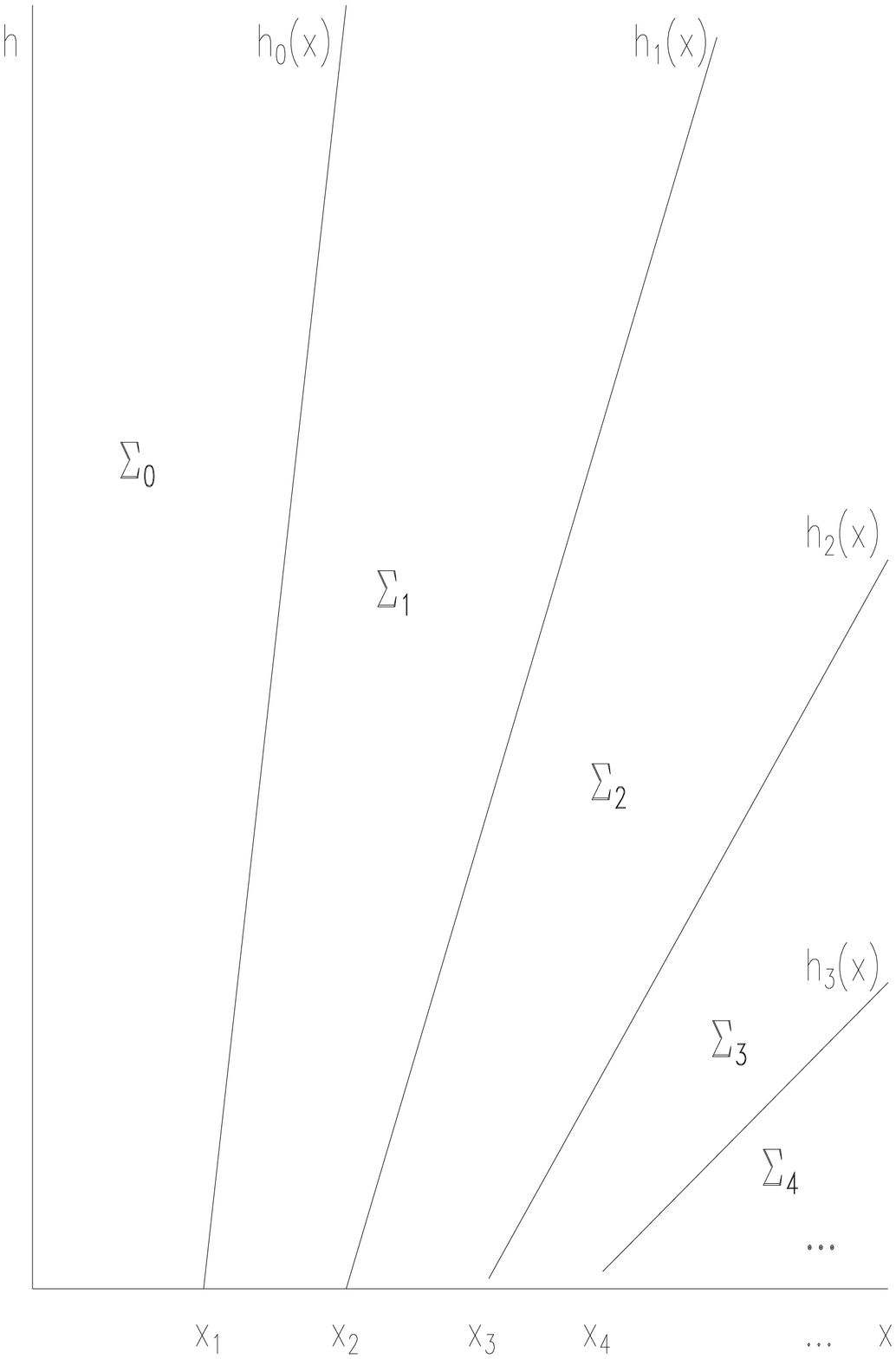}
\end{center}
\end{figure}
\noindent
${\bf 2^\circ.}$ The distortions $x_l$ are different for all the memorized
patterns (4):
$$\vec s^{(l)}=(1,\ldots,1,\underbrace{1-x_l}_l,1,\ldots,1),\quad l=1,2,\ldots,
p.$$
Suppose 
$$x_1\ge x_2\ge\ldots,\ge x_p>0.$$
It can be shown that, firstly, only one of the configuration vectors
$$\vec\sigma^*(k)=(\underbrace{-1,-1,\ldots,-1}_{k},1,\ldots,1),\quad
k=0,1,\ldots,k_{max},$$ 
can be the ground state. Here
$$k_{max} \left\{\begin{array}{cl}=p,&\mbox{ 
when }p<\frac{n}2\\
\le\left[\frac{n}2\right],&\mbox{ when }p>\frac{n}2.
\end{array}\right.$$
And secondly, {\bf in order to the vector $\vec\sigma^*(k)$ be a ground state, the
fulfilment of the inequalities
$$x_{k+1}-\frac{\sum\limits^k_{i=1} x_i -\sum\limits^p_{j=k+2} x_j}{n-2k-1}
\le \ p\ \le x_k-\frac{\sum\limits^{k-1}_{i=1} x_i -\sum\limits^p_{j=k+1}
x_j}{n-2k+1} $$
is necessary and sufficient conditions}. 
\vskip 2mm
\noindent
When $p$ is fixed and $n\to\infty$ these inequalities are much more
simpler:
$$x_{k+1}\le p\le x_k.$$

\noindent
Note, that for sufficiently large $n$ and $p>x_1$, the standard 
$\vec\varepsilon(n)=\vec\sigma^*(0)$ is the ground state. It seems, that for
the Hebb connection matrix the last result clarifies the meaning of the
well-known Latin saying {\it "Repetitio est mater studiorum"} --
showing the same pattern many times (inevitably each time with a distortion),
we seek to make the number of the presentations $p$ greater than the maximal
distortion.

\end{document}